 \documentstyle[12pt]{article}
 \def\beqar {\begin{eqnarray}}
 \def\eeqar {\end{eqnarray}}
 \def\beq {\begin{equation}}
 \def\eeq {\end{equation}}
 \def \ep {{\epsilon}}
 \def \half {{\textstyle{1\over 2}}}
 
 \def \la {{\langle}}
 \def \ra {{\rangle}}
 \def \vf {{\varphi}}
 \def \Tr {{\rm Tr}}
 \def \bz {{\bar z}}
 \def \S {{\cal S}}

 \def \slashint {{\begin{array}{ c }\hskip 
 .016in\lceil\\\rfloor\\\end{array}}}
 \def\O{{\cal O}}
 
 \def\d{\partial}
 \def\bd{{\bar \partial}}
 \def\bra{\langle}
 \def\ket{\rangle}

 \def\bt{\beta}

 \def\al{\alpha}
 
 \def\p{\phi}

 \def\hf{\frac{1}{2}}

 \def\Om{\Omega}

 \def \et {\eta}
 
\begin{document}
 
 \begin{titlepage}
 \null\vspace{-62pt}
 
 \pagestyle{empty}
 \begin{center}
 \rightline{} \rightline{CCNY-HEP-5/02}
 
 \vspace{1.0truein} {\Large\bf Noncommutative gravity: fuzzy sphere 
 and others}\\
 
 \vspace{1in} YASUHIRO ABE and V. P. NAIR\\
 \vskip .1in {\it Physics Department\\ City College of the 
 CUNY\\
 New York, NY 10031}\\
 \vskip .05in {\rm E-mail:
 abe@sci.ccny.cuny.edu\\
 ~~~~~~~~~~~~vpn@sci.ccny.cuny.edu}\\
 \vspace{1.5in}
 
 \centerline{\large\bf Abstract}
 \end{center}
 Gravity on noncommutative analogues of compact spaces can 
 give a
finite mode truncation of ordinary commutative gravity. We 
obtain
the actions for gravity on the noncommutative two-sphere 
and on the noncommutative ${\bf CP}^2$ in 
terms
of finite dimensional $(N\times N)$-matrices.
The commutative large $N$ limit is also discussed.

 \end{titlepage}

 \pagestyle{plain} \setcounter{page}{2} \baselineskip =16pt
 \section{Introduction}
 
 There has recently been a lot of interest in physics on
 nocommutative spaces, partly motivated by the discovery 
 that
 noncommutative spaces can arise as solutions in string and
 $M$-theories \cite{general}. In the matrix model version 
 of
 $M$-theory \cite{matrix}, noncommutative spaces can be 
 obtained as
 $(N\times N)$-matrix configurations whose large $N$-limit 
 will
 give smooth manifolds \cite{taylor}. Fluctuations of 
 branes are
 described by gauge theories and, with this motivation, 
 there has
 recently been  a large number of papers dealing with gauge
 theories, and more generally field theories, on such 
 spaces
 \cite{general}. There is also an earlier line of 
 development,
 motivated by quantum gravity, using the Dirac operator to
 characterize the manifold and using `spectral actions'
 \cite{connes}.
 
 Even apart from their string and $M$-theory connections,
 noncommutative spaces are interesting for other reasons. 
 Many of
 the noncommutative spaces recently discussed have an 
 underlying
 Heisenberg algebra for the different coordinates. A Lie 
 algebra
 structure is more natural from the matrix model point of 
 view;
 these typically lead to noncommutative analogues of 
 compact
 spaces. Because these spaces are described by finite 
 dimensional
 matrices, the number of possible modes for fields on such 
 spaces
 is limited and so one has a natural ultraviolet cutoff. We 
 may
 think of such field theories as a finite mode 
 approximation to
 commutative continuum field theories, providing, in some 
 sense, an
 alternative to lattice gauge theories. Indeed, this point 
 of view
 has been pursued in some recent work \cite{bal2}. While 
 lattice
 gauge theories may be most simply described by standard 
 hypercubic
 lattices, gravity is one case where the noncommutative 
 approach
 can be significantly better. This can provide a 
 regularized
 gravity theory preserving the various desirable 
 symmetries, which
 is hard to do with standard lattice versions. It would be 
 an
 interesting alternative to the Regge calculus, which is
 essentially the only finite-mode-truncation of gravity 
 known with
 the concept of coordinate invariance built in. A
 finite-mode-truncation is not quantum gravity, but it can 
 give a
 formulation of standard gravity where questions can be 
 posed and
 answered in a well defined way.
 
 Partly with this motivation, we have recently suggested a 
 version
 of gravity on noncommutative spaces \cite{nair1}. This led 
 to an
 action for even dimensional, in particular 
 four-dimensional,
 noncommutative spaces generalizing the 
 Chang-MacDowell-Mansouri
 approach used for commutative four-dimensional gravity 
 \cite{CMM}.
 (Some further developments somewhat related to this 
 analysis, as
 well as different approaches, can be traced from reference
 \cite{cham}.) In this paper, we will consider the case of 
 the
 fuzzy sphere in some detail, setting up the required 
 structures,
 eventually obtaining an action for gravity in terms of 
 finite
 dimensional matrices. The large $N$-limit of the action 
 will give
 the usual action for gravitational fields on $S^2$. Some 
 of the
 physically interesting questions related to this 
 framework, such
 as the computation of correlators, are currently under 
 study. 
We also construct the finite dimensional matrix model action
for gravity on noncommutative ${\bf CP}^2$ and indicate
how this may be generalized to ${\bf CP}^n$.
It would also be interesting to extend some of this 
 analysis to
 other two-dimensional noncommutative spaces which have 
 finite
 number of degrees of freedom such as noncommutative 
 Riemann
 surfaces \cite{poly}.

 \section{Derivatives, vectors, etc.}
 
 We shall primarily be concerned with noncommutative (NC) 
 versions
 of coset spaces of the form $G/H$ for some compact Lie 
 group $G$,
 $H$ being a subgroup of $G$. Most of our discussion will 
 be based
 on $S^2 = SU(2)/U(1)$. Functions on the NC sphere are 
 given by
 $(N\times N)$-matrices with elements $f_{mn}$. A 
 particular way of
 representing such functions was given in \cite{nair1} as 
 \beq
 f(g,g') = \la g \vert {\hat f}\vert g' \ra = \sum_{mn} 
 f_{mn}
 {\cal D}^{*(j)}_{mj} (g) {\cal D}^{(j)}_{nj} (g') 
 \label{1} \eeq
 where ${\cal D}^{(j)}_{mk} (g)$ are the Wigner ${\cal
 D}$-functions for $SU(2)$ belonging to the spin-$j$
 representation. Here $N=2j+1$. In this way of representing
 functions, derivatives can be realized as the right 
 translation
 operators $K_a$ on $g$, \beq K_a \cdot {\cal 
 D}^{(j)}_{mk}(g) =
 \left[ {\cal D}^{(j)}\left(g~{t_a}\right) \right]_{mk} 
 \label{2}
 \eeq Here $t_a =\sigma_a/2$, $\sigma_a$  being the Pauli 
 matrices.
 However, to realize various quantities, the action in 
 particular,
 purely in terms of matrices, we will introduce a different 
 but
 related way of defining derivatives, vectors, tensors, 
 etc., on a
 fuzzy coset space.
 
 Let $g$ denote an element of the group $G$ and define \beq 
 S_{Aa}=
 2 ~\Tr (g^{-1} t_A g t_a ) \label{3} \eeq where $t_a$ and 
 $t_A$
 are hermitian matrices forming a basis of the Lie algebra 
 of $G$
 in the fundamental representation. We normalize these by 
 $\Tr (t_a
 t_b)= \half \delta_{ab}$, $\Tr (t_A t_B )= \half 
 \delta_{AB}$.
 (the distinction between upper and lower case indices is 
 only for
 clarity in what follows.) For $SU(2)$, $a, A =1,2,3$ and 
 $S_{Aa}$
 obey the relations \beq
 \begin{array}{r c l}
 S_{Aa} ~S_{Ab} &=&\delta_{ab}\\
 S_{Aa} ~ S_{Ba} &=& \delta_{AB}\\
 \epsilon_{ABC} S_{Aa} S_{Bb} &=& \epsilon_{abc} S_{Cc}\\
 \epsilon_{abc} S_{Aa} S_{Bb} &=& \epsilon_{ABC} S_{Cc}\\
 \end{array}
 \label{4} \eeq
 
 Let $L_A$ be the $(N\times N)$-matrix representation of 
 the
 $SU(2)$  generators, obeying the commutation rules $[L_A, 
 L_B] =
 i \epsilon_{ABC} L_C$. We then define the operators \beq 
 {\cal
 K}_a = S_{Aa} ~L_A ~-~ \half K_a \label{5} \eeq where 
 $K_a$ are
 the right translation operators, $K_a g = g t_a$. One can 
 think of
 them as differential operators, $K_a = i (E^{-1})^i_a (\d 
 / \d
 \vf^i)$, in terms of the group parameters $\vf^i$, $g^{-1} 
 dg =
 (-it_a) E^a_i d\vf^i$. $K_a$ obey the commutation rules 
 $[K_a
 ,K_b]= i \epsilon_{abc} K_c$.  We then find \beq \left[ 
 {\cal K}_a
 , {\cal K}_b \right] = {i\over 4} \epsilon_{abc} ~K_c 
 \label{6}
 \eeq Identifying the $U(1)$ subgroup generated by $t_3$ as 
 the
 $H$-subgroup, we define derivatives on the fuzzy $S^2$  as 
 ${\cal
 K}_\pm = {\cal K}_1 \pm i {\cal K}_2$. Notice that this is 
 a
 hybrid object, being partially a matrix commutator and 
 partially
 something that depends on the continuous variable $g$. 
 This is
 very convenient for our purpose and in the end $g$ will be
 integrated over anyway.
 
 We now define a function $f$ on NC $S^2$ as an $(N\times
 N)$-matrix with no $g$-dependence. The derivative of $f$ 
 is then
 given as \beq {\cal K}_\mu \cdot f \equiv \left[ {\cal 
 K}_\mu , f
 \right] = S_{a\mu} [L_A ,f]\label{7} \eeq $\mu = \pm$. 
 Since
 $\left[ {\cal K}_+ , {\cal K}_- \right] = \half ~ K_3$ 
 from
 (\ref{6}), we find $\left[ {\cal K}_+ , {\cal K}_- \right] 
 \cdot f
 = 0$, consistent with the expectation that derivatives 
 commute
 when acting on a function. Equation (\ref{7}) also shows 
 that it
 is natural to define a vector on NC $S^2$ as \beq V_\mu = 
 S_{A\mu
 }~ V_A \label{8} \eeq where $V_A$ are three $(N\times
 N)$-matrices. On a two-sphere, a vector should only have 
 two
 independent components, so this is one too many and $V_A$ 
 must
 obey a constraint. Notice that the quantity $[L_A, f]$ 
 obeys the
 condition $L_A [L_A,f] + [L_A,f]L_A =0$, since $L_A L_A$ 
 is
 proportional to the identity matrix. This suggests that 
 the
 correct constraint for a general vector is $L_A V_A + V_A 
 L_A =0$.
 In the large $N$-limit, since $L_A$ become proportional to 
 $x_A$,
 the commutative coordinates of the two-sphere as embedded 
 in ${\bf
 R}^3$ (with $x_A x_A = 1$), the condition $x\cdot V=0$ is 
 exactly
 what we need to restrict the vectors to directions 
 tangential to
 the sphere. We may thus regard $L_A V_A + V_A L_A =0$ as 
 the
 appropriate NC version. As we shall see below this 
 constraint will
 also emerge naturally when we define integrals on NC 
 $S^2$. Using
 \beq [K_a , S_{Ab} ] = i \epsilon_{abc} S_{Ac}\label{9} 
 \eeq we
 find \beq \left[ {\cal K}_+ , {\cal K}_- \right] \cdot 
 V_\pm = \pm
 \half~V_\pm\label{10} \eeq which is consistent with the 
 Riemann
 curvature of $S^2$, $R^+_{+-~+} = - R^-_{+--}= \half$. 
 Higher rank
 tensors may also be defined in an analogous way with 
 several
 $S_{Aa}$'s, $ T_{{\mu_1}{\mu_2}\cdots{\mu_r}}= S_{A_1 
 \mu_1}S_{A_2
 \mu_2}\cdots S_{A_r \mu_r}~ T_{A_1 A_2\cdots A_r}$.
 
 We now turn to a definition of `integration' on the  NC 
 $S^2$. We
 will only need, and will only define, integration of the 
 NC
 analogue of an antisymmetric rank-2 tensor or a two-form. 
 Such a
 quantity has components of the form $W_{+-}= (S_{A+} 
 S_{B-}-
 S_{A-} S_{B+} ) W_{AB}$. By the properties of $S_{Aa}$, 
 $S_{A+}
 S_{B-}- S_{A-} S_{B+}= -2i\epsilon_{ABC} S_{C3}$. 
 Integration of
 $W_{+-}$ over $g$ (with the trace of the matrices 
 $W_{AB}$) will
 give zero. To get a nonzero integral we must introduce a 
 density
 factor $\rho$. Such a factor must commute with $K_3$ to be
 properly defined on $SU(2)/U(1)$ and must give nonzero 
 upon
 $g$-integration with $S_{C3}$. The only choice is $\rho = 
 {1\over
 3} S_{K3} L_K$. The appearance of such a density factor is
 actually very natural. If we consider a commutative $S^2$ 
 embedded
 in ${\bf R}^3$ with coordinates $x_A$, then $x_A = 
 S_{A3}$ in a
 suitable parametrization. The usual volume element is 
 oriented
 along $x_A = S_{A3}$ and so we can expect a factor $\rho = 
 {1\over
 3} S_{K3}L_K$ in the NC case. With the introduction of the 
 factor
 $\rho$, we can consider an `integral' of the form $\int_g 
 \Tr
 (\rho W)$. However, if we consider $\int_g \Tr (\rho W f)$ 
 where
 $f$ is a function, we do not have the expected cyclicity 
 property
 since $[\rho ,f]\neq 0$ in general. Cyclicity property can 
 be
 obtained if we symmetrize the factors inside the trace 
 except the
 density factor $\rho$. Gathering these points, we now 
 define an
 `integral' over NC $S^2$, denoted by, $\slashint$, as 
 follows.
 \beq \slashint A_1 A_2 \cdots A_l = \int_g \Tr \left[ \rho
 ~~{1\over l}\sum_{cycl.} (A_1 A_2 \cdots A_l ) \right] 
 \label{11}
 \eeq where $A_1, A_2, \cdots ,A_l$ are functions, vectors,
 tensors, etc., such that the product is an antisymmetric 
 rank-2
 tensor (of the form $W_{+-}$), a NC analogue of a 
 two-form. The
 summation in (\ref{11}) is over cyclic permutations of the
 arguments.
 
 In a similar fashion, let us consider now a NC analogue of 
 an
 exterior derivative, in particular, the analogue of a 
 two-form
 corresponding to the curl of a vector $V_\mu = 
 S_{A\mu}V_A$, $\mu
 =\pm$. Since we have defined ${\cal K}_{\pm}$ as 
 derivatives on NC
 $S^2$, a NC analogue of such a term can be given by \beqar
 dV&\equiv& [{\cal K}_+ ,V_-] - [{\cal K}_- ,V_+] 
 \nonumber\\
 &=& \left( S_{A+}S_{B-} - S_{A-}S_{B+}\right) [L_A,V_B]
   ~-2S_{C3} V_C \nonumber\\
 &=& (-2i) ~S_{C3} \left( \ep_{ABC} [L_A,V_B] -i V_C 
 \right)
 \label{12} \eeqar If $h$ is a function on NC $S^2$, we 
 also have
 \beqar V ~dh &\equiv& V_+ [{\cal K}_- ,h] - V_-
 [{\cal K}_+ ,h] \nonumber\\
 &=& (-2i) \ep_{ABC} S_{C3} V_A [L_B ,h]\label{13} \eeqar 
 Using the
 definition of the integral (\ref{11}) we find \beqar 
 \slashint
 dV~h &=& (-2i) {1\over 2} \Tr \biggl[ L_K \left\{ 
 \ep_{ABC}
 [L_A,V_B] -i V_C\right\}~h +\left\{ \ep_{ABC} [L_A,V_B] -i
 V_C\right\} L_K
 ~h\biggr] \nonumber\\
 &&\hskip 1.5in \times\int_g {1\over 3} S_{K3}S_{C3}
 \nonumber\\
 &=& (-2i) {1\over 2}\Tr \biggl[ L_C \left\{ \ep_{ABC} 
 [L_A,V_B] -i
 V_C\right\}~h +\left\{ \ep_{ABC} [L_A,V_B] -i V_C\right\} 
 L_C
 ~h\biggr]
 \nonumber\\
 \label{14} \eeqar where we used $\int_g S_{K3} S_{C3} =
 3\delta_{KC}$. Similarly we have \beq \slashint V~dh = 
 (-2i)
 {1\over 2}\Tr \biggl[ \ep_{ABC} (L_C V_A +V_A L_C) 
 [L_B,h]\biggr]
 \label{15} \eeq By using cyclicity of the trace for the 
 finite
 dimensional matrices $L_A, V_B, h$, etc., we find that the 
 desired
 partial integration property \beq \slashint dV~h = 
 \slashint V~dh
 \label{16} \eeq holds if $V_A$ obey the constraint \beq 
 L_A ~V_A
 +V_A~ L_A =0\label{17} \eeq 
This relation, which was 
 introduced
 earlier, is now seen, based on integration properties, to 
 be the
 correct constraint for vectors.
When $V_A$ are gauge fields, this constraint will have to be slightly
modified for reasons of gauge invariance, as will be seen in the next
section.
 
 \section{Action for gravity on NC $S^2$}
 
 We are now in a position to discuss actions for gravity on 
 NC
 $S^2$. As outlined in \cite{nair1}, we introduce the gauge 
 field
 \beqar
 {\cal A}_\mu &=& {\cal A}_\mu^A I^A = e^a_\mu I^a +
 \Omega^3_\mu I^3 +\Omega^0_\mu I^0\nonumber\\
&=& e^+_\mu I^+ +e^-_\mu I^- +\Omega^3_\mu I^3 +\Omega^0_\mu I^0
\label{18}
\eeqar
The  components
 $(\Omega^0_\mu , ~\Omega^3_\mu,~e^a_\mu )$ are vectors on 
 NC $S^2$
 as defined in the previous section. The upper indices of 
 these
 vectors correspond to components for the Lie algebra of 
 $U(2)$,
 $(I^0, I^3, I^a)$, $a=\pm$, form the $(2\times 
 2)$-representation
 of $U(2)$. Specifically, in terms of the Pauli matrices
 $\sigma_i$, $I^0 =\hf {\bf 1}$, $I^3 =\hf \sigma_3$, 
 $I^\pm = \hf
 ( \sigma_1 \pm i \sigma_2)$. ${\cal A}_\mu$ is thus a 
 vector on NC
 $S^2$ which also takes values in the Lie algebra of 
 $U(2)$. This
 $U(2)$ is the group acting on the upper indices of ${\cal 
 A}_\mu$
 or the tangent frame indices. Notice that, with $L_A$, 
 $K_a$ and
 the $I$'s, we have three different actions for $SU(2)$. In 
 terms
 of ${\cal A}_\mu$ we now define \beq [{\cal K}_\mu + {\cal 
 A}_\mu
 , {\cal K}_\nu + {\cal A}_\nu ] = {i\over 4} 
 \ep_{\mu\nu\alpha}
 K_\alpha ~+~ F_{\mu\nu} \label{19} \eeq
 
 In our description, gravity is parametrized in terms of 
 deviations
 from $S^2$. The vectors $e^a_\mu$ ($a=\pm$) are the frame 
 fields
 for this and $\Om_\mu^{\al}$ ($\al=0,3$) are the spin 
 connections.
 As opposed to the commutative case, there can in general 
 be a
 connection for the $I^0$ component, since we need the full 
 $U(2)$
 to form NC gauge fields. One can expand $F_{\mu\nu}$ as 
 \beq
 F_{\mu\nu}= F_{\mu\nu}^0 ~I^0 ~+~ {\cal R}_{\mu\nu}^3 ~I^3 
 ~+~
 {\cal T}_{\mu\nu}^a ~I^a \label{20} \eeq 
${\cal T}_{\mu\nu}^a$ is
 the torsion tensor. On commutative $S^2$, ${\cal 
 R}_{\mu\nu}^3$ is
 of the form $R_{\mu\nu}(\Omega ) + 2 (e^+_\mu e^-_\nu - 
 e^-_\mu
 e^+_\nu )$ where $R_{\mu\nu}(\Omega )$ is the Riemann 
 tensor. For
 NC $S^2$, the expression for ${\cal R}_{\mu\nu}^3$ is a 
 little
 more involved.
 
 In defining an action, we will use our prescription for the integral.
The gauging of ${\cal K}_\mu$ is equivalent to
the gauging $L_a \rightarrow L_A + {\cal A}_A$. Thus we must also 
change our definition of $\rho$  to  $\rho ={1\over 3} S^K_3 (L_K
+{\cal A}_K)$. The constraint (\ref{17}) is now replaced by
\beq
(L_A +{\cal A}_A) (L_A +{\cal A}_A) = L_A L_A
\label{20a}
\eeq
(${\cal A}_A$ is expanded in terms of the $I^a$ as in (\ref{18}).)
This equation was first proposed in \cite{kar} as the correct condition
to be used for gauge fields.

The
data for gravity is presented in the form of a gauge 
 field and
 the action suggested in \cite{nair1}, generalizing the
 McDowell-Mansouri approach for commutative gravity, is 
 \beq {\cal
 S}= \alpha \slashint {\rm tr} ( Q F ) \label{21} \eeq 
Here 
 ${\rm
 tr}$ denotes the trace over the $I$'s regarded as 
 $(2\times
 2)$-matrices. For higher even dimensional cases it would 
 be
 $\slashint{\rm tr}(QFF...F)$, where $Q$ is a combination 
 of the
 $I$'s which commutes with the $H$-subgroup of $G$, 
 $G=U(2)$ for
 the present case; for NC $S^2$, $Q =I^3$. However, unlike 
 the case
 of four and higher dimensions, $\slashint{\rm tr}(I^3 F)$
 vanishes, which is the NC analogue of the statement that 
 the
 two-dimensional Einstein-Hilbert action $\int R \sqrt{g}$ 
 is a
 topological invariant. As noted in the commutative 
 context, we
 must use a Lagrange multiplier scalar field $\eta$ to 
 obtain
 nontrivial actions. In the present case, the analogous 
 action is
 \beq \S  = \al \slashint {\rm tr} (I^3 \eta F ) \label{22} 
 \eeq
 Here $\eta = \eta^0 I^0 + \eta^3 I^3 + \eta^+ I^+ +
\eta^- I^-$, 
 $(\eta^0
 ,\eta^3, \eta^a)$ being scalar functions on NC $S^2$. 
 Using the
 decomposition (\ref{20}) for the field strength, we can 
 simplify
 this expression as 
\beq
 \S =  -i{\alpha \over 2}\Tr \biggl[ I^3 \eta \left[(L_C+{\cal A}_C)
F_C + F_C (L_C+{\cal A}_C)\right]\biggr]
\label{22a} 
\eeq 
where $F_C = F_C^0 I^0 + F_C^3 I^3 + F_C^+ I^+ + F_C^- I^-$, 
\beqar 
F^0_C&=& 
 \hf
 \left\{[L_A,\Om^0_B] + \hf (\Om^0_A \Om^0_B + \Om^3_A 
 \Om^3_B)
 +(e^{+}_A e^{-}_B + e^{-}_A e^{+}_B)\right\}\epsilon_{ABC}
-{i\over 2} \Omega^0_C\label{23} \\
 F^3_C&=& \hf \left\{  [L_A,\Om^3_B] + \hf (\Om^0_A \Om^3_B 
 +\Om^3_A \Om^0_B) +(e^{+}_A e^{-}_B - e^{-}_A 
 e^{+}_B)\right\}
 \epsilon_{ABC} -{i\over 2} \Omega^3_C
 \label{24} \\
 F^-_C&=& \hf \left\{[L_A,e^-_B] + \hf e^-_A (\Om^0_B + 
 \Om^3_B )
 -\hf (\Om^0_B-\Om^3_B) e^-_A\right\} \epsilon_{ABC} -{i\over 2}e^-_C
 \label{25} \\
 F^+_C&=&\hf \left\{[L_A,e^+_B] + \hf e^+_A (\Om^0_B - 
 \Om^3_B )
 -\hf (\Om^0_B +\Om^3_B) e^+_A\right\}\epsilon_{ABC} -{i\over 2} e^+_C 
  \label{26}
 \eeqar

 Equation ({\ref{22a}), with (\ref{23}-\ref {26}), is the 
 action
 for gravity on the NC $S^2$. They are expressed entirely 
 in terms
 of finite dimensional $(N\times N)$-matrices, $L_A, 
 e^{a}_A$, and
 $\Om_A^{\al}$. As mentioned earlier, this action can give 
 us an
 alternative to the Regge calculus, and, in principle, we 
 can
 calculate many interesting physical quantities, 
 correlation
 functions in particular, from (\ref{22a})-(\ref{26}) by 
 analyzing
 it as a matrix model. In this paper, however, we shall not
 calculate correlators. Instead, we shall analyze the 
 action a bit
 further and discuss the commutative limit in what follows.
 
 Variations of the action with respect to $\et$'s then 
 provide the
 four equations of motion, 
\beq 
{\cal F}^a \equiv \left[(L_C+{\cal A}_C)
F_C + F_C (L_C+{\cal A}_C)\right]^a =0\label{27} 
\eeq 
for $a = 0,~3, \pm$. The 
 components $a
 =\pm$ correspond to the vanishing of torsion. ${\cal F}^3$ 
 is not
 quite the Riemann tensor associated with $\Om^3$, due to 
 the
 $e^+e^-$-term. The vanishing of ${\cal F}^3$ shows that 
 the
 Riemann tensor is proportional to the $e^+e^-$-term.
 
 There are also equations of motion associated with the 
 variation
 of the $e^\pm ,~\Om^3 ,~\Om^0$, which are coupled 
 equations for
 the four $\eta$'s. We will not write them out here, they 
 can be
 easily worked out from the expressions 
 (\ref{23}-{\ref{26}) for
 the $F_C$'s. Notice however that one solution of the 
 equations of
 motion is easy to find. The variation of the action with 
 respect
 to the $e^\pm ,~\Om^3 ,~\Om^0$ is of the form 
\beq 
 \delta\S =
 -i{\alpha \over 2} \Tr \biggl[ I^3 \eta ~
\delta \left[(L_C+{\cal A}_C)
F_C + F_C (L_C+{\cal A}_C)\right] \biggr]\label{28} 
\eeq Thus
$\eta =0$ is evidently a solution.
 
 The equations for the connections $e^\pm ,~\Om^3 ,~\Om^0$ 
 in
 (\ref{27}) are also solved by setting all $F_{\mu\nu}$ to 
 zero.
 This corresponds to the choice ${\cal A}_\mu = S_{B\mu} 
 {\cal
 A}_B$, ${\cal A}_B = i U^{-1} [L_B, U]$ where $U$ is a 
 matrix
 which is an element of $U(N)\otimes U(2)$, $L_B$ is viewed 
 as $L^A
 \otimes {\bf 1}$. In other words, it is an element of 
 $U(2)$ with
 parameters which are $(N\times N)$-matrices. This 
 corresponds to
 the NC $S^2$ itself.
 
 \section{Commutative limit}
 
 We now consider the commutative limit of the action 
 obtained in
 the previous section; this can be realized by taking a 
 large $N$
 limit. In this section, we will take the dimension of 
 matrices
 $L_A$'s as $N+1$, rather than $N$. The large $N$-limit can 
 be
 taken easily by writing the $L_A$'s as follows 
 \cite{nair2}.
 \beqar L_+ =& \frac{N+2}{2} \p_+  + z^2 \d_z \qquad& \p_+ 
 =
 \frac{2z}{1+z\bz}
 \nonumber \\
 L_- =& \frac{N+2}{2} \p_- - \d_z      \qquad& \p_- =
 \frac{2\bz}{1+z\bz}
 \nonumber \\
 L_3 =& \frac{N+2}{2} \p_3 + z\d_z     \qquad& \p_3 =
 \frac{1-z\bz}{1+z\bz} \label{30} \eeqar A basis of states 
 on which
 $L_A$ act is given by $|\al \ket$, $\al=0,1,\cdots,N$ with 
 $\bra z
 | \al \ket = 1, z, \cdots, z^N$. If we choose such a basis 
 the
 vectors, $\Om$ and $e$, can be considered as functions of 
 $z,
 \bz$. The commutative limit can then be taken by the 
 following
 replacements. \beqar
 L_A & \longrightarrow &  \frac{N+2}{2} \p_A \nonumber \\
 \left[L_A, \Om_B\right] & \longrightarrow &  \frac{1}{N+1}
 \left\{\frac{N+2}{2}\p_A , \Om_B \right\} \nonumber \\
 && = \frac{N+2}{2}\frac{1}{N+1}(1+z\bz)^2(\d \p_A \bd 
 \Om_B
 -\bd \p_A \d \Om_B) \nonumber \\
 && = \frac{N+2}{2} \frac{1}{N+1}k_A \Om_B \label{31} 
 \eeqar where
 $\d=\frac{\d}{\d z}$, $\bd=\frac{\d}{\d \bz}$ and the 
 operators
 $k_A$ are given in the appendix. The replacement of a 
 commutator
 with a Poisson bracket is analogous to the passage from 
 the
 quantum theory to the classical theory, $1/(N+1)$ serving 
 as the
 analogue of Planck's constant. 
Notice also that, in the expression
$L_A +{\cal A}_A$, the term $L_A$ dominates in the large
$N$ limit. The formulae
given above 
 and the
 passage to the large $N$-limit can be best seen by 
 geometrical
 quantization of $S^2$.
 
 It is instructive to consider the large $N$-limit of one 
 of the
 terms in the action, say the term involving
$\eta^0$ in some detail. 
 Denoting this term as
$\S [\eta^0]$  and using
 (\ref{31}), we find 
\beqar \S[\et^0] &=& -i\frac{\al}{2} 
 \left(
 \frac{N+2}{2}\right) ~\ep_{ABC} \Tr ~ \left[ \et^0 \left[
 {N+2\over 2(N+1)}~\phi_Ck_A\Om^3_B + \phi_C (e^+_A e^-_B - 
 e^-_A
 e^+_B)    \right] + \O \left(\frac{1}{N}\right)
 \right]\nonumber \\
 &\approx& -i \bt\frac{1}{N+1}~\ep_{ABC} \Tr ~  \left[ 
 \et^0
 \phi_C\left[ {1\over 2}~k_A\Om^3_B +  (e^+_A e^-_B - e^-_A 
 e^+_B)
 \right] + \O \left(\frac{1}{N}\right) \right] \label{32} 
 \eeqar
 Here $\bt = \frac{\al}{2} \left( \frac{N+2}{2}\right) 
 (N+1)$. This
 will be taken as an $N$-independent constant. In carrying 
 out
 these simplifications, it is useful to keep in mind that 
 the
 $\Om_A$ obey the constraint \beq \p_A~\Om_A +\Om_A~\p_A 
 \approx 2
 \phi_A \Om_A \approx 0 \label{38} \eeq which is a natural
 reduction of the NC constraint in (\ref{17}). Noting that 
 $\hf
 k_A$ can be defined as a derivative operator on $S^2$, we 
 can
 express \beq k_A \Om_B \equiv 2~D_A \Om_B \label{39} \eeq 
 Also the
 trace over matrices can be replaced by the integral over 
 $z$ and
 $\bz$ according to \beq \frac{1}{N+1}\Tr \longrightarrow 
 \int
 \frac{dz d\bz} {\pi (1+ z \bz)^2} \equiv \int_{z, \bz} 
  \label{40}
 \eeq We can now rewrite (\ref{32}) as \beq 
 \S[\et^0]\approx -i \bt
 ~ \ep_{ABC} \int_{z, \bz} \et^0 \phi_C \left( D_A \Om^3_B 
 + (e^+_A
 e^-_B - e^-_A e^+_B) \right) \label{41} \eeq 
Similar 
 results can
 be obtained for the rest of $\et$'s. With a simple 
 arrangement of
 notation, (\ref{41}) and the analogous formulae for the 
 other
 $\eta$'s, we recover the commuatative action 
\beq \S \sim
 \ep_{AB}\int_{z, \bz} \et F_{AB} \label{42} 
\eeq 
This is 
 the
 two-dimensional Jackiw-Teitelboim action on $S^2$ 
 \cite{jackiw}.
 We have therefore checked that, in the large $N$ limit, 
 the
 gravitational action (\ref{22a}) on NC $S^2$ does reduce 
 to the
 commutative one.

\section{Generalizations}

Eventhough we have derived the matrix action (\ref{22a}) via
our defintions ${\cal K}_\mu$, the final result is simple
and can be interpreted more directly. The key quantity that enters
in the action is the combination $L_A + {\cal A}_A^a I^a$.
We can write this as
\beqar
L_A + {\cal A}_A^a I^a &=& D_A^a I^a~\equiv~D_A \nonumber\\
D_A^0 &=& 2 L_A + {\cal A}_A^0 \nonumber\\
D_A^a &=& {\cal A}_A^a, \hskip 1in a\neq 0\label{43}
\eeqar
The key ingredient is thus a set of $(N\times N)$ hermitian 
matrices
$D_A^a$. The definition of the curvatures is seen to be
\beqar
\bigl[ D_A, D_B \bigr]&=& \bigl[ D_A^a I^a , D_B ^b I^b \bigr] = i
\epsilon_{ABC}  D_C^cI^c ~+~ F_{AB}^cI^c\nonumber\\
&=&  i \epsilon_{ABC}  D_C ~+~ F_{AB}
\label{44}
\eeqar
The action is given by 
\beqar
{\cal S} &=& -i {\alpha \over 2}  \Tr \biggl[
I^3\eta ~ \epsilon^{ABC}\left( D_C F_{AB} + F_{AB}D_C\right)
\biggr]\nonumber\\
&=& -2i\alpha \Tr \biggl[ I^3\eta~ \left( \epsilon^{ABC} D_A D_B D_C
- i D^2\right) \biggr] 
\label{45}
\eeqar
The constraint on the the $D$'s is $D_A D_A = L_A L_A$.
It is only in this constraint that the restriction to the sphere
arises. Notice that
for this particular case, we could absorb
the factor of $I^3$ inside the trace into the
field $\eta$.

The general structure is as thus follows. We start with 
an irreducible finite dimensional
representation of the Lie algebra of $SU(2)\times U(1)$ 
given by the $I^a$
with the commutation relation $[I^a, I^b ]= i f^{abc} I^c$.
(Specifically, here $f^{abc} =\epsilon^{abc}$ for $a,b,c = 3, \pm$
and zero otherwise.) We then construct the combinations
$D_A = D_A^a I^a$ where the $D_A^a$ are hermitian arbitrary
matrices of some given dimension $N$.  Using the
same $SU(2)$ structure constants we define the
curvatures by $F_{AB} = [D_A , D_B] - i f_{ABC} D_C$.
This does not make any reference
to the sphere yet. We restrict to the sphere by imposing
the constraint $D_A D_A = L_A L_A$.
The action is then constructed in terms of $F_{AB}$ as
in (\ref{45}). 

We can use this to generalize to $SU(3)$, which will
apply to the case of gravity on noncommutative
${\bf CP}^2$. Let $I^a$, $a=1, 2,...,8$ be a set
of $(3\times 3)$- matrices forming
a basis of the
Lie algebra of $SU(3)$,
with the commutation rules
$[I^a ,I^b ]=i f^{abc} I^c$. We include
$I^0 = {1\over \sqrt{6}} {\bf 1}$ to
make up the algebra of $U(3)$. Let $L_A$ denote
an irreducible
representation of the $SU(3)$ algebra in terms of
$(N\times N)$-matrices, with $[L_A , L_B ]= if_{ABC}L_C$.
The dynamical variables are then given by
$D_A^a$ which are 
a set of arbitrary $(N\times N)$-matrices.
(There are 72 matrices since $A=1,2,...,8$ and
$a= 0,1,...,8$.) The curvatures are defined by
$F_{AB} =[ D_A , D_B] -if_{ABC} D_C $,
$D_A = D_A^a I^a$.
As the constraints to be obeyed by the $D$'s we choose
\beqar
D_A D_A &=& L_A L_A\nonumber\\
d_{ABC} D_B D_C &=& c~~ D_A\label{46}
\eeqar
where $c$ is some constant and $d_{abc} = \Tr (I_a I_b I_c 
+I_a I_c I_b)$ upto a constant normalization
factor. The continuum limit of these
conditions are well known as giving ${\bf CP}^2$ as an algebraic surface
in
${\bf R}^8$. It has also been noted that they can give 
noncommutative ${\bf CP}^2$ \cite{seif}.
Following the construction of the action given in \cite{nair1}
and our general discussion in section 3,
we can write the action for
gravity on
${\bf CP}^2$ as
\beqar
\S &=& \alpha ~\Tr \biggl[ I^8\left(  D_A F_{KL} F_{MN}+
 F_{KL} F_{MN} D_A\right) \biggr] f_{KLB} f_{MNC}  d_{ABC}\nonumber\\
&=& \alpha ~\Tr \Biggl[ I^8\biggl(  D_A \{[D_K, D_L]-if_{KLR}D_R\} 
\{[D_M, D_N]-if_{MNS}D_S\} \nonumber\\
&&\hskip .5in +
\{[D_K, D_L]-if_{KLR}D_R\}\{[D_M, D_N]-if_{MNS}D_S\} D_A\biggr) \Biggr]
f_{KLB} f_{MNC}  d_{ABC}\nonumber\\
\label{47}
\eeqar
This action, along with the constraints (\ref{46}), gives
gravity on noncommutative ${\bf CP}^2$ as a matrix model.
One can also check directly that the large $N$ limit of this will
reduce to the MacDowell-Mansouri version of the
action for gravity on
commutative
${\bf CP}^2$.

It is clear that similar actions can be constructed
for all ${\bf CP}^n$. Notice that the
quantity
$f_{KLB} f_{MNC}  d_{ABC}$ is the fifth rank invariant of $SU(3)$. 
For ${\bf CP}^n$ we can use $n$ factors of
$F$'s and one factor of $D$ and then contract indices with
$\omega^{A_1 ...A_{2n+1}}$, the
invariant tensor  of $SU(n+1)$ with rank $(2n+1)$.
Thus
\beq
\S = \alpha \Tr \Biggl[ I^{((n+1)^2- 1)}~\biggl( 
D_{A_1}F_{A_2 A_3} F_{A_4 A_5}...+ F_{A_2 A_3} F_{A_4 A_5}...D_{A_1}
\biggr)\Biggr] \omega^{A_1 ...A_{2n+1}}\label{48}
\eeq
In the large $N$ limit, such an action will contain the Einstein term
(in the MacDowell-Mansouri form), but will also have terms with higher
powers of the curvature. The action (\ref{48}) has to be supplemented
by suitable constraints on the $D$'s, which may also be 
taken as (\ref{46}) \cite{bal4}.

\section*{Acknowledgements}
 
 This work was supported in part by NSF grant number 
 PHY-0070883
 and by a PSC-CUNY grant.
\hfil\break
{\it Note added}: In a recent paper \cite{bal3},
the analogue of the Euler number for NC $S^2$ is defined using
Ginsparg-Wilson type operators. The density for this index 
reduces to the form $(L_A +A_A) F_{BC}\epsilon_{ABC}$
in the large $N$ limit, similar to what we have used in place of the
curvature in the action ({\ref{22a}). However, in \cite{bal3},
only a $U(1)$ type field is considered, while we have a $U(2)$
field which can lead to the torsion-free condition
as an equation of motion. We thank Professor Balachandran
for discussions on this point.

 \section*{Appendix}
 The operators $k_A$ in section 4 are defined as follows.
 $$
 \left\{\p_A , \Om_B \right\} = (1+z\bz)^2(\d \p_A \bd 
 \Om_B -\bd
 \p_A \d \Om_B) \equiv k_A \Om_B
 $$
 \vspace{-0.7cm} \beqar
 k_+ &=& 2( z^2 \d + \bd)     \nonumber \\
 k_- &=& -2( \d + \bz^2 \bd)\nonumber \\
 k_3 &=& 2( z \d - \bz \bd)\nonumber \eeqar
 These satisfy the following properties. \beqar
 k_+\p_+ =0      \qquad & k_+\p_-=4\p_3  \qquad & 
 k_+\p_3=-2\p_+ \nonumber \\
 k_-\p_+ =-4\p_3 \qquad & k_-\p_-=0      \qquad & 
 k_-\p_3=2\p_- \nonumber \\
 k_3\p_+ = 2\p_+ \qquad & k_3\p_-=-2\p_- \qquad & k_3\p_3=0
 \nonumber \eeqar \vspace{-0.5cm}
 $$
 \left[\frac{k_+}{2},\frac{k_-}{2}\right]=2\frac{k_3}{2}, 
 \quad
 \left[\frac{k_3}{2},\frac{k_+}{2}\right]=\frac{k_+}{2},\quad
 \left[\frac{k_3}{2},\frac{k_-}{2}\right]=-\frac{k_-}{2}
 $$
 Notice that $\hf k_A$ form an $SU(2)$ algebra.

 \end{document}